\documentclass[sigconf,natbib=false]{acmart}

\usepackage{graphicx}
\usepackage[style=ACM-Reference-Format,backend=bibtex,sorting=none]{biblatex}
\usepackage[ruled,linesnumbered]{algorithm2e}
\usepackage{booktabs}
\usepackage{multirow}
\usepackage{bm}
\newcommand{\argmin}{\operatornamewithlimits{argmin}}
\newcommand{\argTopk}{\operatornamewithlimits{argTopk}}

\AtBeginDocument{%
  \providecommand\BibTeX{{%
    \normalfont B\kern-0.5em{\scshape i\kern-0.25em b}\kern-0.8em\TeX}}}

\setcopyright{acmcopyright}
\copyrightyear{2018}
\acmYear{2018}
\acmDOI{XXXXXXX.XXXXXXX}

\acmConference[Conference acronym 'XX]{Make sure to enter the correct
  conference title from your rights confirmation emai}{June 03--05,
  2018}{Woodstock, NY}
\acmPrice{15.00}
\acmISBN{978-1-4503-XXXX-X/18/06}

\begin{document}

\title{Adaptive Domain Interest Network for Multi-domain Recommendation}
\author{Yuchen Jiang, Qi Li, Han Zhu, Jinbei Yu, Jin Li, Ziru Xu, Huihui Dong, Bo Zheng}
\email{{jiangyuchen.jyc, luyuan.lq, zhuhan.zh, jinbei.yjb, echo.lj, ziru.xzr, dhh267344, bozheng}@alibaba-inc.com}
\affiliation{%
  \institution{Alibaba Group}
  \country{China}
}

\begin{abstract}
Industrial recommender systems usually hold data from multiple business scenarios and are expected to provide recommendation services for these scenarios simultaneously. In the retrieval step, the topK high-quality items selected from a large number of corpus usually need to be various for multiple scenarios. 
Take Alibaba display advertising system for example, not only because the behavior patterns of Taobao users are diverse, but also differentiated scenarios' bid prices assigned by advertisers vary significantly. 
Traditional methods either train models for each scenario separately, ignoring the cross-domain overlapping of user groups and items, or simply mix all samples and maintain a shared model which makes it difficult to capture significant diversities between scenarios. In this paper, we present \textbf{A}daptive \textbf{D}omain \textbf{I}nterest network that adaptively handles the commonalities and diversities across scenarios, making full use of multi-scenarios data during training. Then the proposed method is able to improve the performance of each business domain by giving various topK candidates for different scenarios during online inference. Specifically, our proposed ADI models the commonalities and diversities for different domains by shared networks and domain-specific networks, respectively. In addition, we apply the domain-specific batch normalization and design the domain interest adaptation layer for feature-level domain adaptation. A self training strategy is also incorporated to capture label-level connections across domains.
ADI has been deployed in the display advertising system of Alibaba, and obtains 1.8\% improvement on advertising revenue.

\end{abstract}

\ccsdesc[500]{Information systems~Information retrieval~Retrieval models and ranking}

\keywords{Recommender Systems, Information Retrieval, Neural Networks, Multi-domain Learning}


\maketitle


\section{Introduction}



Modern recommendation systems and advertising systems are usually built as a pipeline, including retrieving, ranking, reranking and other parts. For retrieval step~(also known as matching step), the most important objective is to retrieve topK high-quality items from a very large corpus~(millions) for downstream ranking task in limited time. Item-CF~\cite{sarwar2001item} and user-CF~\cite{han2011data} are the most lightweight and common methods used in retrieval, which leverage user/item collaborative signals. However, with the proliferation of deep learning methods, deep models for recommendation perform better than algorithms based solely on collaborative signals. YouTube product-DNN~\cite{covington2016deep,yi2019sampling} proposes to generate user/item vector representations, calculating their inner product and then retrieving items using efficient approximate k-nearest neighbor searches. Facebook EBR~\cite{huang2020embedding} integrate embedding-based retrieval with boolean matching retrieval in their search engines to address the semantic matching issues. Further works focus on either the capability of vector representations or searching strategies for retrieval. MIND~\cite{li2019multi} proposes a multi-interest retrieve model using a dynamic routing mechanism. In the meantime, TDMs~\cite{zhu2018learning,  zhu2019joint, zhuo2020learning} and Deep-Retrieval~\cite{gao2020deep} are proposed to increase model complexity 
by building the searching index for the large corpus. Although the deep models are thriving in recommendation system, traditional recommenders and online advertising system mainly focus on how to model single scenario well. In this paper, we are devoted to get benefit from multiple scenarios' data in the retrieval stage.


Data collected from multiple business scenarios own commonalities and diversities. For the former, there is an overlap between both users and items for different domains, in other words, the domain-invariant user interest and item information can be transferred from one domain to another. Taking Alibaba display advertising production data for example, there are 49\% of users and 79\% of items appear at least two scenarios based on our data analysis on traffic logs in a day. For the latter, data distributions from multiple domains are different, since users' preference, items' displaying permission and advertiser's bidding price are quite various across domains. To address those issues, three solutions are generally adopted in real world systems. The first solution is to train one model per domain using training data collected from this domain, but the drawbacks~\cite{sheng2021one} are obvious: 1). Maintaining multiple models for multiple domains need much human operation cost and calculating resources. 2). Separately training models per domain makes it impossible to exploit domain-invariant knowledge, especially for those minor business scenarios where training data is limited.
A step further solution is to mix data and train a shared model. By doing this, human operation cost and calculating resources can be saved. But if without particular design, model performance may decrease when different domains conflict.
The last generally adopted solution is to train a unified model in a multi-task manner. Although the multi-domain recommendation task usually shares the similar model structure with multi-task recommendation task, we argue that those two tasks are fundamentally different.
For the former, data distributions of inputs from multiple domains are quite different but the the task goals are the same~(such as multi-domain semantic segmentation in ~\cite{zou2019confidence}). For the latter, labels from different task vary significantly~(such as CTR and CVR prediction in ~\cite{ma2018entire}) while the input is the same. Therefore, a particular architecture designed for multi-domain recommendation is needed in real world applications.

Existing efforts for multi-domain recommendation ~\cite{sheng2021one, shen2021sar} focus on ranking step, while retrieval step is rarely studied ~\cite{hao2021adversarial}. To solve the multi-domain recommendation in retrieval step, we come up with the \textbf{A}daptive \textbf{D}omain \textbf{I}nterst network, which learns users' preferences for multiple scenarios simultaneously. Firstly, our proposed ADI models the commonalities and diversities for different domains by common networks and domain-specific networks, respectively. To tackle the feature-level domain adaptation, we present two domain adaptation methods, which are domain-specific batch normalization and domain interest adaptation layer. In addition, to capture label-level connection across domains, a self training method is also incorporated.
 
To summarize, our proposed method achieves following contributions:
\begin{itemize}
\item We proposed a novel model architecture named \textbf{A}daptive \textbf{D}omain \textbf{I}nterest network to tackle multi-domain recommendation in retrieval step. The ADI network efficiently learn the commonalities and diversities for multiple domains, leading to an overall performance lift for all domains.
\item We provide domain interest adaptation layers to dynamically transfer raw input features to domain-related features. Extensive experiments and visualization prove the effectiveness of the proposed domain interest adaptation layer.
\item We get a first attempt to apply self training method on our multi-domain recommendation problem, capturing the potential label-level cross-domain connection.
\item We conduct solid experiments on real-world industry production dataset and deploy our proposed method in the online advertising system.
\end{itemize}

\section{Related Works}



\textbf{Domain Adaptation:}
Domain Adaptation $\left(DA\right)$ problem is a branch of transfer learning, aiming to learn from source domain then get better performance on target domain. The key to solving the DA problem lies in transferring useful knowledge from source domain to target domain. 
A lot of works~\cite{ganin2015unsupervised, tzeng2017adversarial} extract domain-invariant features by minimizing the cross-domain difference of feature distributions through adversarial learning. DSN~\cite{bousmalis2016domain} designs domain separation network to transfer knowledge through the shared-network. Some works~\cite{inoue2018cross, zou2018unsupervised} handle DA problem through a self-training/pseudo-labeling strategy. Previously DA methods mostly focus on solving problems in computer vision and neural language process, 
while recently there are more and more researches on recommendation systems especially on CTR prediction. 

\noindent\textbf{Multi-Domain Recommendation:}Multi-domain recommendation  ~\cite{man2017cross, li2020ddtcdr} task aims to improve model performance on each  domain using knowledge transferred from the other domains. The difference between multi-domain recommendation$\left(MDR\right)$ ~\cite{sheng2021one, shen2021sar, hao2021adversarial} and cross-domain recommendation$\left(CDR\right)$ ~\cite{li2020ddtcdr, hu2018conet} lies in transfer directions. CDR aims to transfer knowledge in a specific direction (for example, using  data from main scenario to improve performance in cold-start scenario), while MDR is aiming to gain an overall performance in all domains.
SAR-Net~\cite{shen2021sar} accommodates the transfer of users' interest across scenarios through two specific attention modules, and uses mixture of experts to extract the required information. STAR~\cite{sheng2021one} proposes star topology consisting of shared centered parameters and domain-specific parameters, to keep one model get refined CTR prediction for different domain. AFT~\cite{hao2021adversarial} 
proposes a novel adversarial learning method to solve the MDR problem.

\noindent\textbf{Multi-Task Learning:}
Multi-task learning ~\cite{ruder2017overview} is a machine learning paradigm to learn several related tasks at the same time, leading to a better performance on each task. There are many general multi-task model structures, which gain significant progress in computer vision~\cite{heuer2021multitask}, neural language processing ~\cite{liu2019multi}, information retrieval~\cite{jain2021mural} and recommendation system~\cite{ma2018entire}. Share-Bottom~\cite{caruana1997multitask} network designs a shared network at the bottom to learn the similarity and multiple task-specific network at the top to learn the differences. Multi-gate Mixture-of-Experts (MMoE)~\cite{ma2018modeling} use multiple expert networks at the bottom to capture different patterns in the data and learn a per-task and per-sample weighting of each expert networks allowing different tasks to utilize experts differently. However, the MMoE architecture meets worse performance when tasks correlation is complex, which is called \textbf{\textsl{seesaw phenomenon}}. PLE~\cite{tang2020progressive} eases above issue by separating experts network into the task-related private network and task-independent shared network. SNR-Net~\cite{ma2019snr} learns a optimized combination of several shared sub-networks.

\noindent\textbf{Self Training:}
Self training~(or self-supervised learning) is a learning strategy, in which labeled data are limited and much more unlabeled data are available. Exploiting unlabeled data is the key to gaining a performance boost in self training paradigm.
~\cite{zou2018unsupervised, zou2019confidence} apply self training method on DA problems by producing pseudo-labels for the target domain.
For multi-domain recommendation task, most work focus on sample level~\cite{wang2019minimax}, feature level~\cite{shen2021sar}, and parameter level~\cite{sheng2021one} transferring to gain a better performance. However, we argue potential connection in the labels from different domains can be employed by our proposed efficient self training method, which has been neglected by existing works~\cite{sheng2021one, shen2021sar}.

\begin{figure*}[htbp]
 \setlength{\abovecaptionskip}{1pt}
  \includegraphics[width=\textwidth]{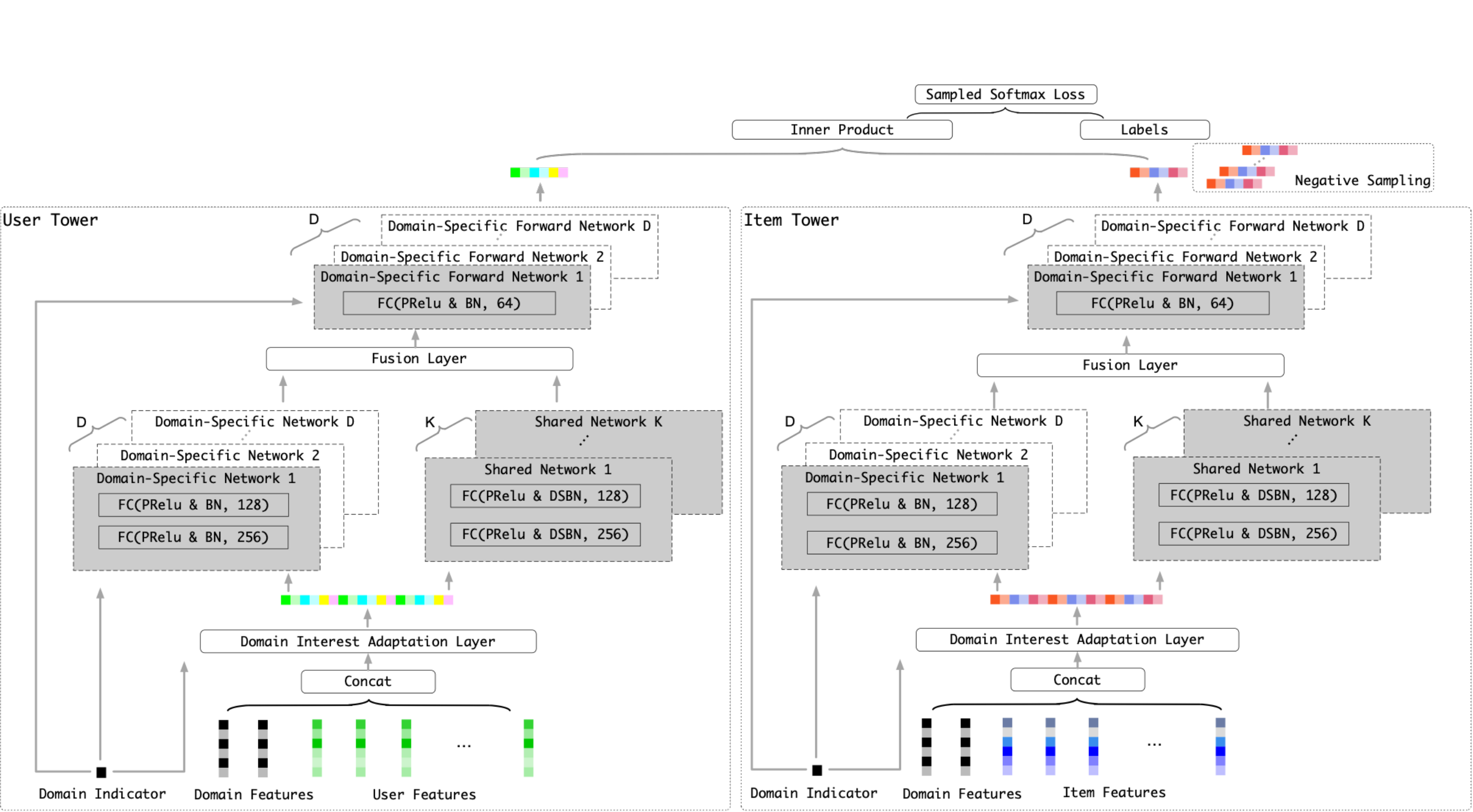}
  \caption{An illustration of the overall architecture of ADI. Following the gray arrow, a sample will be firstly embeded, then fed into the Domain Interest Adaptation Layer, Shared \/ Domain-Specific Network, Fusion Layer and Domain-Specific Forward Network. After getting user/item representations through user/item tower, the inner product will be produced and the sampled softmax loss will be calculated in the end. The domain indicator is ultilized for choosing which domain-related network to use.}
  \label{fig:arch}
\end{figure*}

\section{Preliminaries}
\subsection{Problem Formulation}
In this section, we formalize the definition of multi-domain retrieval task. Multi-domain retrieval task aims to retrieve high-quality items for multiple domains from a very large corpus. More specifically, let $\mathcal{U}$ and $\mathcal{V}$ denote the user set and the item set, respectively. The online multi-domain retrieval task can be formulated as follows:



\begin{equation}\label{eq:domain-topk}
    \mathcal{S}_{u,d} = \argTopk_{v \in \mathcal{V}} f_\theta(v|u,d),
\end{equation}
where $d$ denotes the domain indicator, and $f_{\theta}(v|u,d)$ is the estimated matching function with trainable parameters $\theta$ for measuring the quality of $u$ to $\mathcal{V}$ given the user $u$ and domain indicator $d$. $\mathcal{S}_{u,d}$ is a set containing topK items with respect to $f_{\theta}(v|{u,d})$. 

In neural based retrieval models, learning such a model $f_\theta(v|u,d)$ can be regarded as an instance-level classification problem. Distribution of positive item $v$ win from $\mathcal{V}$ is based on the softmax function, i.e.,
\begin{equation}\label{eq:softmax}
    s_{\theta}(v|u,d) = \frac{\exp(f_{\theta}(v|u,d))}{\sum_{v' \in \mathcal{V}} \exp(f_{\theta}(v'|u,d))}. 
\end{equation}
Then $\theta$ is trained to minimize the negative log likelihood $-\log s_{\theta}(v|u,d)$ over the training data 
\begin{equation}\label{eq:obj}
    \theta^* = \argmin_{\theta} \sum_{d} \sum_{u} \sum_{v \in \mathcal{B}_{u,d}} - \log s_{\theta}(v|u,d),
\end{equation}
where $\mathcal{B}_{u,d}$ is the set of interacted items by $u$ given the user $u$ and domain indicator $d$.

In practice, since $\mathcal{V}$ is usually extremely large, sub-sampling is widely adopted to reduce the computational complexity of computing the denominator of Eq. (\ref{eq:softmax}) .
Following~~\cite{covington2016deep,li2019multi}, we use the sampled softmax loss~~\cite{jean2015using} and replace $f_{\theta}(v|u,d)$ in Eq. (\ref{eq:softmax}) with 
\begin{equation}\label{eq:sampledsoftmax}
    \tilde{f}_{\theta}(v|u,d)=f_{\theta}(v|u,d)-\log Q(v).
\end{equation}
With the sub-sampling, we have Eq. (\ref{eq:objnew}). $\mathcal{N}_{u,d}$ is the set of irrelevant items, which are sampled from $\mathcal{V}$ according to the proposal distribution $Q: \mathcal{V} \to \mathbb{R}$ such that its size satisfies $|\mathcal{N}_{u,d}| \ll |\mathcal{V}|$.
{\footnotesize
\begin{equation}\label{eq:objnew}
    \theta^* = \argmin_{\theta} \sum_{d,u,v \in \mathcal{B}_{u,d}} - \tilde{f}_{\theta}(v|u,d) + \log \left( \exp (\tilde{f}_{\theta}(v|u,d)) + \sum_{v' \in \mathcal{N}_{u,t}} \exp (\tilde{f}_{\theta}(v'|u,t)) \right).
\end{equation}
}


\section{Methodology}
In this section, we introduce our proposed method to tackle multi-domain retrieval problem. The overall model architecture are shown in Figure \ref{fig:arch}. The total model architecture is  designed to commonalities and diversities for different domains from three angles. Firstly, the backbone network extracts parameter-level commonalities and diversities from data collected from different domains. Secondly, the domain adaptation methods learn feature-level diversities. Lastly, the self-training strategy captures label-level commonalities.

\begin{table}[htb]
    \centering
     \caption{A brief feature description for Alibaba display advertising, consisting of user features, item features, and domain features.}
    \label{tab:features}
    \resizebox{0.5\textwidth}{!}{
    \begin{tabular}{l | l}
    \toprule
        Feature Type &  Feature Description\\
    \hline
        User Features & User profiles, User behaviors~(click, add\_to\_cart, pay..), etc. \\
        Item Features & Item attributes, Creative attributes, Advertiser attributes, etc. \\
        User Domain Features & User behaviors by domain, etc.\\
        Item Domain Features &  Bidding Price by domain, Item statistics by domain~(ctr, click..), etc.\\
        Domain Indicator Feature & Domain indicator \\
    \bottomrule
    \end{tabular}
    }
\end{table}

\subsection{Backbone Network}
To efficiently learn the commonalities and diversities between data distributions from different domains, we design the shared networks and the domain-specific networks at the bottom with domain-specific forward networks at the top. Such architecture is able to perform better when dealing with multi-domain retrieval problems compared to vanilla DNN~\cite{covington2016deep}, share-bottom network~\cite{caruana1997multitask} and MMoE~\cite{ma2018modeling}, which is proved in the following experiments part.
\subsubsection{Shared Embedding Layer}
As shown in Table~\ref{tab:features}, the training/testing samples contain rich feature information. Therefore, the first step is to transfer those high dimensional sparse one-hot vectors into low-dimensional embedding vectors, and all domains share the same embedding layers.
\begin{equation}
    F_i = EMBED(f_i),
\end{equation}
\begin{equation}\label{eq:user_embedding}
    {\bf{F}} = concat(F_1~|\cdot\cdot\cdot|~F_n),
\end{equation}
where $F_i$ denotes  $i_{th}$ embeded feature. $\bf{F}$ denotes user/item inputs. 
\subsubsection{Shared Network \& Domain-Specific Network}
After obtaining encoded user representations and item representations, we introduce the shared network and domain-specific network as shown in Figue ~\ref{fig:arch}. Inspired by~\cite{bousmalis2016domain}, we design the shared network to learn representations shared by all domains and the domain-specific network to learn domain-specific representations in each domain:
\begin{equation}\label{eq:share_gate}
    \alpha_k = \frac{W_{shared}^k(f_{domain})+b_{shared}^k}{\sum_{n=1}^K(W_{shared}^n(f_{domain})+b_{shared}^n)},
\end{equation}
\begin{equation}\label{eq:share_net}
    {E_{shared}} =  \sum_{k=1}^K\alpha_kMLP_{shared}^k({\bf{F}}),
\end{equation}
\begin{equation}\label{eq:spec_net}
    {E_{spec}^{(d)}} = MLP_{spec}^{(d)}({\bf{F}}^{(d)}),
\end{equation}
where $MLP$ denotes the multilayer perceptron, $f_{domain}$, ${\textbf{F}}^{(d)}$ denote domain-related features and data collected from domain $d$, respectively. In our practice, we use domain indicator embedding as $f_{domain}$. $W_{shared}^n, b_{shared}$ are weights and bias of a one-layer shallow neural network. Data from all domains will feed into shared networks, while data from domain $d$ will be feed into $d_{th}$ domain-specific network. More specifically, suppose there are training data from $D$ domains, we will build $K$ shared network and $D$ specific network. The total number of FCs is $D+K$

\subsubsection{Fusion Layer}\label{sec:fusion}
The fusion layer aims to learn an optimized combination of outputs from Domain-Specific Network and Shared Network, which can be described as follows:
\begin{equation}\label{eq:fusion}
    \beta_1^{(d)} = \sigma(W_{fusion\_spec}^{(d)}(f_{domain})),
\end{equation}
\begin{equation}
    \beta_2^{(d)} = \sigma(W_{fusion\_shared}^{(d)}(f_{domain})),
\end{equation}
\begin{equation}\label{eq:mul_fusion}\small
    {E_{fusion}^{(d)}} = concat(\beta^{(d)}_1 E_{spec}^{(d)}~|~\beta^{(d)}_1E_{spec}^{(d)}\odot{\beta^{(d)}_2E_{shared}}~|~\beta^{(d)}_2E_{shared});
\end{equation}
where $\sigma$ denotes sigmoid function, $\odot$ denotes hadamard product, and $\beta_1^{(d)}, \beta_2^{(d)}$ denote feature weights assigned for $E_{spec}^{(d)}, E_{shared}$, respectively. We name the proposed fusion layer the \textbf{CONCAT} version. Therefore, shared and specific network will produce domain-related $E_{fusion}^{(d)}$ for each domain. Besides, we implement two variants, which are \textbf{SUM} version used by MMoE~\cite{ma2018modeling}, SAR-Net~\cite{shen2021sar} and \textbf{Network-Mul} version proposed by STAR~\cite{sheng2021one}.
For the \textbf{SUM} version, we use the gating network of MMoE as the fusion layer. $W_{gate}, b_{gate}$ denote weights and bias of the gating network:
\begin{equation}\label{eq:fusion}
    \alpha^{(d)} = \sigma(W_{gate}^{(d)}(f_{domain})+b_{gate}),
\end{equation}
\begin{equation}\label{eq:fusion}
    {E_{fusion}^{(d)}} = {\alpha^{(d)}}E_{spec}^{(d)} + {(1-\alpha^{(d)})}E_{shared},
\end{equation}
For the \textbf{Network-Mul} version, we use the STAR-Topology FCN of STAR~\cite{sheng2021one} as the fusion layer~(Note that STAR only has one shared network), $W_{shared}$, $b_{shared}$ and $W_{spec}^{(d)}$, $b_{spec}^{(d)}$ denote parameters in $FC_{shared}$ and $FC_{spec}$, respectively:
\begin{equation}\label{eq:mul_fc}
    FC_{Net-Mul}(X) = (W_{shared}\odot{W_{spec}^{(d)}})\cdot{\textbf{X}} + b_{shared} + b_{spec}^{(d)},
\end{equation}
\begin{equation}\label{eq:fusion}
    {E_{fusion}^{(d)}} = FC_{Net-Mul}({{\bf{F}}^{(d)}}),
\end{equation}
Experiments in Section~\ref{sec:fusion_experiment} prove our proposed \textbf{CONCAT} version achieves the best performance, which is adopted as the fusion layer.
\subsubsection{Domain-Specific Forward Network}
After obtaining domain-related $E_{fusion}^{(d)}$, finally, the outputs will feed into domain-related forward network, which describes as follows:
\begin{equation}\label{eq:tower}
    {E} = FC_{forward}^{(d)}(E_{fusion}^{(d)}).
\end{equation}
the output $E$ produced by user tower and item tower will be used for following inner product and sampled softmax calculating.

\subsection{Domain Adaptation}
We provide two approaches to solving domain adaptation problems in the multi-domain  recommendation task: {\bf{domain-specific batch normalization}} and {\bf{domain interest adaptation layer}}.

\begin{figure}[htbp]
 \setlength{\abovecaptionskip}{2pt}
 \setlength{\belowcaptionskip}{1pt}
  \includegraphics[width=0.5\textwidth]{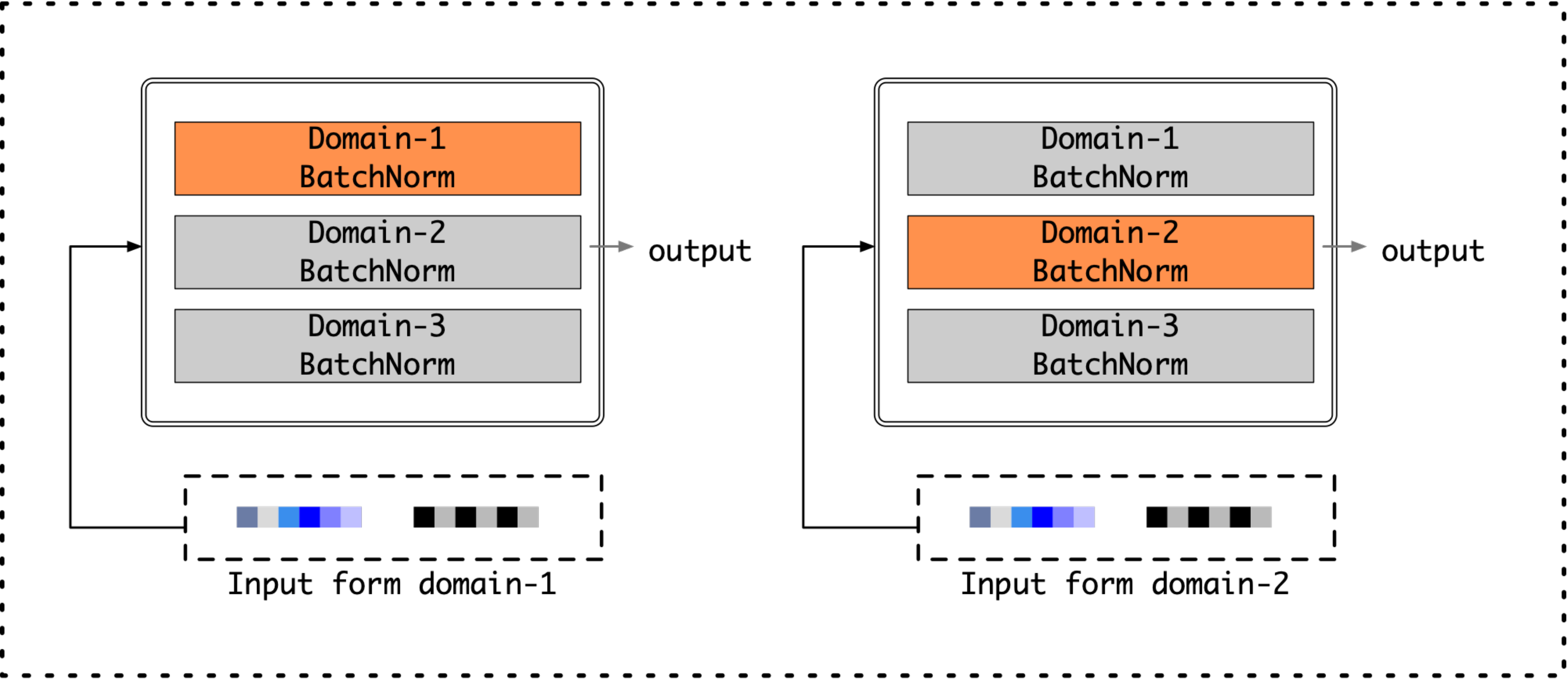}
  \caption{An illustration of DSBN. Samples collected from different domains will choose different branches of DSBN.}
  \label{fig:dsbn}
\end{figure}
\subsubsection{Domain-Specific Batch Normalization}
Batch normalization technology~(BN) ~\cite{ioffe2015batch} has been widely used to train very deep neural network. Let $\mu$ denotes the mean value of input ${\bf{X}}$, while $\sigma^2$ denotes the variance. The batch normalization method can be described as follows:
\begin{equation}\label{eq:bn}
    \hat{{\bf{X}}} = \alpha\frac{{\bf{X}}-\mu}{\sqrt{\sigma^2+\epsilon}}+\beta,
\end{equation}
where $\alpha$ and $\beta$ are learnable parameters, $\epsilon$ is a very small quantity to avoid the denominator being 0. BN assumes that the input ${\bf{X}}$ satisfies the assumption of independent and identical distribution(i.i.d), which works well in single scenario. However, multi-domain retrieval problem is facing with a mixed data distributions. Calculating the global BN parameters and ignoring statistic discrepancies between different domains may hurt the final performance. Inspired by~\cite{chang2019domain}, We apply the domain-specific batch normalization~(DSBN) to solve the mentioned problem:
\begin{equation}\label{eq:bn}
    \hat{{\bf{X}}}^{(d)} = \alpha^{(d)}\frac{{\bf{X}}^{(d)}-\mu^{(d)}}{\sqrt{(\sigma^{(d)})^2+\epsilon}}+\beta^{(d)},
\end{equation}
where ${\bf{X}}^{(d)} \in \bf{X}$ denotes collected samples from domain $d$. By estimating domain-specific batch statistics $\mu^{(d)},(\sigma^{(d)})^2,\alpha^{(d)},\beta^{(d)}$ in batch normalization, we trust that the model is able to to capture the domain-specific information.

\subsubsection{Domain Interest Adaptation Layer}
The domain interest adaptation layer comes from the intuition that different domains are supposed to focus on different parts of the raw features. We implement three types of domain interest adaptation layer: linear domain transformation, vanilla domain attention, and SE-Block based domain attention:\\
\noindent{\bf{Linear domain transformation:}}
Linear domain transformation used by~\cite{shen2021sar} maps original features into domain-related features. Let $F_i^{(d)}$ denotes $i_{th}$ feature of embedded input collected from domain $d$, and $N$ is total feature number. $W^{(d)}, b^{(d)}$ share the same dimension with input $F^{(d)}$. Linear domain transformation method describes as follows:
\begin{equation}\label{eq:lt_concat}
    F^{(d)} = concat(F_1^{(d)}~|~\cdot\cdot\cdot~|~F_N^{(d)}),
\end{equation}
\begin{equation}\label{eq:lt}
    \hat{F}^{(d)} = W^{(d)}\odot F^{(d)}+b^{(d)},
\end{equation}

\noindent{\bf{Vanilla domain attention:}}
Let $Q^{(d)}_i$ denotes $i_{th}$ domain-specific query vector for attention weights calculating and $\alpha_i^{(d)}$ denotes the attention weight of $i_{th}$ feature, the vanilla domain attention mechanism describes as follows:
\begin{equation}\label{eq:att_alpha}
    \alpha^{(d)}_i =\sigma(Q^{(d)}_i{F^{(d)}_i}),
\end{equation}
\begin{equation}\label{eq:vatt}
    \hat{F}^{(d)} = concat(\alpha_1^{(d)}F_1^{(d)}~|~\cdot\cdot\cdot~|~\alpha_n^{(d)}F_N^{(d)}),
\end{equation}
{\bf{SE-Block based domain attention:}}
Squeeze-and-Excitation Network~(SE-Net)~\cite{hu2018squeeze} has achieved SOTA results in many computer vision tasks~\cite{wang2019towards}. We argue that SE-Block is another form of attention mechanism, capturing the difference in the importance of features in different domains. $F_{se}$ denotes a $(FC, Relu, FC)$ block and $F_{avg}$ denotes average pooling operator. ${\bm{\alpha}}^{(d)}$ denotes $N-dimensional$ SE attention scores vector for domain $d$.
\begin{equation}\label{eq:se_alpha}
    {\bm{\alpha}}^{(d)} = F_{se}(concat(F_{avg}({F_1^{(d)}})~|~\cdot\cdot\cdot~|~F_{avg}(F_N^{(d)}))),
\end{equation}
\begin{equation}\label{eq:vatt}
    \hat{F}^{(d)} = {\bm{\alpha}}^{(d)}\odot{concat(F_1^{(d)}~|~\cdot\cdot\cdot~|~F_N^{(d)})}.
\end{equation}
SE-Block based domain adaptation layer learns different domain attention weights for different domains, transferring cross-domain knowledge in a lightweight and efficient way.

By adding domain interest adaptation layer into the backbone network, the raw features are transferred into domain-related features. Experiments and visualization in Section~\ref{sec:discussion} prove the effectiveness of the proposed domain interest adaptation layer.

\subsection{Self Training}
\begin{algorithm}
\caption{Self Training For Multi-Domain Retrieval}\label{algorithm}
\SetKwInOut{Input}{input}\SetKwInOut{Output}{output}
\KwIn{Matching function $f_\theta(X,d)$, training data $(X^{(d)}, d)$ sampled from domain $d$ with label $y^{(d)}$, hyper-parameter $p$ to determine the portion of pseudo-labels.}
\KwOut{The trained parameter $\theta^*$.}
Initializing $\theta$ randomly. \\
\For{$epoch\leftarrow 1$ \KwTo $EPOCHS$}{
    \For{$iter\leftarrow 1$ \KwTo $ITERATION\_NUM$}{
        \For{$i\leftarrow 1$ \KwTo $D$}{
        Calculate loss for all training data $(X^{(i)}, y^{(i)})$\\
            \For{$j\leftarrow 1$ \KwTo $D$}{
                \If{$i\neq j$}
                    {Embed $(X^{(i)}, i)$ sampled from domain $i$\\
                    Compute prediction scores for $(X^{(i)}, j)$ \\
                    Sort by $Scores^{(j)}_{(i)} = f_\theta(X^{(i)},j)$ \\
                    Select top $p$ percent samples as pseudo data $(X_p^{(i)}, y_p^{(j)})$\\
                    Calculate loss for pseudo data $(X_p^{(i)}, y_p^{(j)})$
                }
            }
        }
        $p = p + \Delta{p}$\\
    }
}
\end{algorithm}

\begin{table*}[htbp]
\caption{Overall performance comparisons on Alibaba production dataset, we use Recall@N as the metric .}\vspace{5pt}
\label{tab:overall_table}
\centering
\resizebox{0.8\textwidth}{!}{
\begin{tabular}{l l c c c c c c c c}
\hline \hline
\multirow{2}*{Method} & \multicolumn{3}{c}{ Domain \#1} & \multicolumn{3}{c}{ Domain \#2} & \multicolumn{3}{c}{ Domain \#3}\\
\cmidrule(lr){2-4} \cmidrule(lr){5-7} \cmidrule(lr){8-10}& R@100 & R@500 & R@1000 & R@100 & R@500 & R@1000 & R@100 & R@500 & R@1000 \\
\hline
DNN-Single    & 0.1862 & 0.3934 & 0.5109 & 0.1170 & 0.2653 & 0.3647 & 0.1029 & 0.2401 & 0.3189 \\
DNN           & 0.1845 & 0.3877 & 0.5077 & 0.1548 & 0.3354 & 0.4429 & 0.1168 & 0.2606 & 0.3597 \\ 
Shared-Bottom & 0.1418 & 0.3330 & 0.4323 & 0.1247 & 0.2980 & 0.3890 & 0.0796 & 0.2196 & 0.2860 \\
Cross-Stitch  & 0.1476 & 0.3245 & 0.4416 & 0.1306 & 0.2901 & 0.3978 & 0.0879 & 0.2054 & 0.3028 \\
MMoE          & 0.1860 & 0.3988 & 0.5142 & 0.1633 & 0.3549 & 0.4619 & 0.1097 & 0.2609 & 0.3530 \\
PLE & \underline{0.1908} & \underline{0.4076} & \underline{0.5248} & \underline{0.1777} & \underline{0.3799} & \underline{0.4901} & \underline{0.1102} & \underline{0.2693} & \underline{0.3662} \\
\hline
ADI-LT       & 0.1935 & 0.4086 & 0.5244 & 0.1691 & 0.3627 & 0.4694 & 0.1018 & 0.2483 & 0.3387 \\
ADI-VA       & 0.2062 & 0.4292 & 0.5465 & 0.1865 & 0.3905 & 0.5005 & 0.1123 & 0.2712 & 0.3672 \\
ADI-SE       & 0.2325 & 0.4683 & 0.5874 & 0.2064 & 0.4222 & 0.5337 & 0.1259 & 0.2982 & 0.3999 \\
ADI-SE~(MAX) & \textbf{0.2438} & \textbf{0.4849} & \textbf{0.6039} & \textbf{0.2219} & \textbf{0.4429} & \textbf{0.5547} & \textbf{0.1468} & \textbf{0.3292} & \textbf{0.4342} \\
\hline
\hline
\end{tabular}}
\end{table*}
Self training methods~\cite{zou2018unsupervised, zou2019confidence, } have been proved as an efficient learning strategy for exploiting unlabeled data during model training. We apply this technology on the multi-domain recommendation in retrieval step for two reasons: 1).~There is a potential label-level connection in training data when there are data overlaps between domains. To be more specific, a interacted item by a user in one domain may still be interacted by the same user in another domain. This assumption works especially when larger domain helps minor domains or even new domains where labeled data are limited. 2).~Adding pseudo-labeled data into training inevitable change the original data distribution, however, we argue that our proposed self training method is more suitable for retrieval models rather than ranking models. Ranking models in advertising systems need to predict precise CTR scores~\cite{zhou2018deep}, adding extra pseudo-labeled data may lead to unknowable performance since the data distribution has been changed and CTR models are sensitive to the data distribution. However, retrieval models for advertising systems aim to provide candidates set for downstream tasks. In other words, precise CTR score are not necessary for retrieval models since multiple candidates will be generated equally. Therefore additional potential interest signals can be added into the model even if the data distribution is slightly changed for generating high-quality topK candidates. Existing methods mostly focus on sample-level~\cite{wang2019minimax}, feature level~\cite{shen2021sar}, and parameter level~\cite{sheng2021one} transferring, while neglecting label-level transferring. Therefore, we proposed this efficient self training method to mine the potential label-level transferring knowledge though domains, which has been proved effective in our experiments.


Given an item $v$ interacted by user $u$ in domain $d$, the self training method follows two steps. \textbf{a)}.~freeze the model to generate pseudo-labels for $v$ in other domains except domain $d$. \textbf{b)}.~freeze the pseudo-labels then fine-tune the model. Following Algorithm~\ref{algorithm}, for each step, we select pseudo-labels with highest confidence scores, and the selection portion gradually increases during training. After obtaining pseudo-labels for $v$ in other domains, $\theta$ in Eq.\ref{eq:obj} is trained to minimize the negative log likelihood $-\log s_{\theta}(v|u,d)$ over the training data and pseudo-labeled data:
\begin{equation}\label{eq:st_loss}
    \theta^* = \argmin_{\theta} \sum_{d} \sum_{u} \sum_{v \in \mathcal{B}_{u,d}} - (\log s_{\theta}(v|u,d) + \log s_{\theta}(\tilde{v}|u,d)).
\end{equation}
where $\tilde{v}$ is the selected potential positive pseudo-items given user $u$ and domain $d$.


\begin{table}[htbp]
 \setlength{\abovecaptionskip}{0.5pt}
 \setlength{\belowcaptionskip}{0.5pt}
\caption{Overall performance comparisons on WSDM Cross-Market dataset. We use Recall@10 as metric and our ADI achieves average SOTA performance.}\vspace{5pt}
\label{tab:overall_table_pub}
\centering
\resizebox{0.4\textwidth}{!}{
\begin{tabular}{l c c c c}
\hline \hline
{Method} & { Domain \#1} &{ Domain \#2} & { Domain \#3} & { Avg} \\
\hline
DNN-Single    & 0.250  & 0.535  & \textbf{0.572} & 0.453\\
DNN           & 0.193  & 0.380  & 0.460  & 0.344   \\ 
Shared-Bottom & 0.261  & 0.507  & 0.537 & 0.435   \\
Cross-Stitch  & 0.252  & 0.404  & 0.510 & 0.388  \\
MMoE          & \textbf{0.273} & 0.611  & 0.368  & 0.417  \\
PLE           & 0.201  & 0.548 & 0.525  & 0.424  \\
ADI           & 0.266  & \textbf{0.614} & 0.519 & \textbf{0.466}  \\
\hline
\hline
\end{tabular}}
\end{table}

\section{Experiments}
\subsection{Experimental Settings}
\subsubsection{Dataset Description}
\begin{table}[h]
 \setlength{\abovecaptionskip}{0.5pt}
 \setlength{\belowcaptionskip}{0.5pt}
    \centering
    \caption{Statistics of datasets used in experiments.}
    \label{tab:dataset}
    \scalebox{0.65}{
    \begin{tabular}{c c c c c c c}
    \toprule
        Dataset  & Users  & Items & Record in D\#1 & Record in D\#2 & Record in D\#3 \\
    \midrule
        Alibaba Production & 92,664,693 & 2,240,723 & 1,364,983,432 & 167,494,696 & 609,611,432\\
        WSDM Cross-Market & 16,903 & 10,994 & 77,173 & 48,302 & 23,367\\
    \bottomrule
    \end{tabular}
    }
\end{table}

\begin{table}[h]
 \setlength{\abovecaptionskip}{0.5pt}
 \setlength{\belowcaptionskip}{0.5pt}
    \centering
    \caption{User/Item Overlap between different domains.}
    \label{tab:dataset_overlap}
        \scalebox{0.65}{
        \begin{tabular}{|c | c | c c c|}
        \toprule
        Dataset & User/Item & D\#1 & D\#2 & D\#3 \\
        \midrule
        \multirow{3}*{\shortstack{Alibaba\\Prodcution}} &
        D\#1 & 78,336,157/1,735,326   & 6,840,732/1,295,028 & 7,043,405/1,525,081\\
        & D\#2 & - &  11,147,803/1,383,050  & 5,551,052/1,279,533\\
        & D\#3 & - & - &  13,099,654/1,998,209 \\
        \midrule
        \multirow{3}*{\shortstack{WSDM\\Cross-Market}} &
        D\#1 & 6466/9762   & 0/1190 & 0/752\\
        & D\#2 & - &  7109/2198  & 0/816\\
        & D\#3 & - & - &  3328/1245 \\
        \bottomrule
        \end{tabular}
    }
  
\end{table}

Two datasets are used to validate our proposed ADI. One is real industrial data named Alibaba Display Advertising Data, and the other is publicly accessible data called WSDM Cross-Market Recommendation Data. Table ~\ref{tab:dataset} and Table ~\ref{tab:dataset_overlap} show the statistic information of this two datasets. The full description are described as follows.

\noindent\textbf{Alibaba Display Advertising Data:}
The Alibaba production data regarding advertising exposure to the consumer on 3 business domains as positive sample, which is collected from traffic logs of the Alibaba online display advertising system and divided by scenarios' bid price assigned rules. The dataset consists of billions of samples with user behavior, ad attribute and ad domain-specific statistic features.

\noindent\textbf{WSDM Cross-Market Recommendation Data:}
The user purchase and rating data on various markets with a considerable number of shared item subsets, provided in "WSDM 2022 CUP - Cross-Market Recommendation". We regard 5 ratings samples as positive. The training dataset consists of millions of examples with no other features but userId and itemId. So the feature-level domain attention method cannot be evaluate on this dataset. The original dataset contains three source domains and two target domains. To keep a consistent model structure with the former, without loss of generality, we use only three source domains to evaluate our proposed method.

\subsubsection{Comparing methods}
\begin{itemize}
\item {\textbf{DNN-Single}}: This method is a implement of YouTube DNN~\cite{yi2019sampling}, one of the most well-known recommendation method in industry. In this version, we train $D$ models for $D$ domains separately. All comparing methods train one model for all domains except this one.
\item {\textbf{DNN}}: This version is YouTube DNN~ trained with mixed data from different domains.
\item {\textbf{Shared-Bottum}~\cite{caruana1997multitask}}: This method is a classical architecture for multi-task/domain adaptation, in which all tasks share one common network at the bottom and each task utilize separated network at the top.
\item {\textbf{Cross-Stitch}~\cite{misra2016cross}}: Cross-Stitch designs $D$ networks at the bottom and learns a optimized linear combination of bottom-network outputs as its task-specific outputs.
\item {\textbf{MMoE}~\cite{ma2018modeling}}: MMoE designs $N$ expert networks at the bottom to capture task-related signals and learns a combination of $N$ expert networks trough its gating mechanism.
\item {\textbf{PLE}~\cite{tang2020progressive}}: PLE is optimized version of MMoE. By designing task-specific expert and common expert to alleviate the seesaw phenomenon. 
\item {\textbf{ADI-LT}}: Our first variant version of ADI, equipped with DSBN, linear transformation and self-training module.
\item {\textbf{ADI-VA}}: Our second variant version of ADI, equipped with DSBN, vanilla domain attention and self-training module.
\item {\textbf{ADI-SE/ADI-SE~(MAX)}}: Our third variant version of ADI, equipped with DSBN, SE-Block based domain adaptation layer and self-training module. The difference between ADI-SE and ADI-SE~(MAX) is that ADI-SE~(MAX) contains more than one shared network in the backbone network.
\end{itemize}
\subsubsection{Evaluation and Metrics}
For Alibaba Display Advertising dataset, data of one day from 3 business domains are used for training and the data of the following day is used for testing. For WSDM Cross-Market Recommendation dataset, testing data is given along with training data, in which each user has 100 candidate items in contrast to retrieval from whole corpus and only one of them is the positive sample. Following existing works~\cite{zhu2018learning}, we use Recall@N as our performance metrics.

\subsubsection{implementation details}
All comparing methods share same input and equipped with BN (ADI is equipped with DSBN). Besides, for a fair comparison, the sizes of the model parameters are kept the same for all methods. For example, suppose there are 3 domains, we set 3 domain-specific networks and 1 shared network for PLE and ADI. At the meantime we set 4 expert networks for MMoE. Each network share exactly the same model size.

\begin{figure*}[htbp]
 \setlength{\abovecaptionskip}{0.5pt}
 \setlength{\belowcaptionskip}{0.5pt}
  \includegraphics[width=\textwidth]{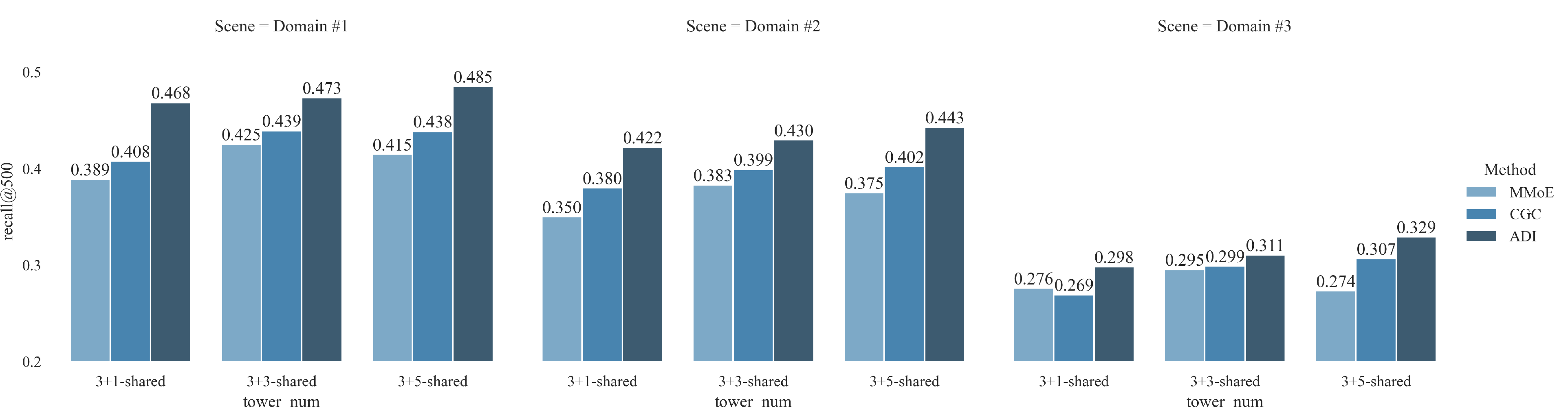}
  \caption{The impact on numbers of shared networks. (3+K shared) in x-axis denotes 3 domain-specific and K shared networks}
  \label{fig:expert_num}
\end{figure*}

\subsection{Overall Performance}
\noindent\textbf{Alibaba Production:} As shown in table.~\ref{tab:overall_table}, our method outperforms among existing works significantly. It is worth noting that the performance of DNN on domain \#2, domain \#3 is better than DNN-single, while performance on domain \#1 is worse. Such phenomenon shows that simply mix training data without designing particular model architecture may hurt the model performance.

Different variants shows the impact of different domain interest adaptation layers. ADI-LT meets the worst performance, while ADI-SE achieves the best performance. 
To sum up, ADI-VA and ADI-SE both achieve overall performance uplift for all domains.The recall of ADI-SE~(MAX)~ is better than ADI-SE proves that using more shared networks uplifts the model performance, but the model complexity also increases. It is a trade-off of choosing a proper model with good performance and affordable parameter complexity.

\noindent\textbf{WSDM Cross-Market:} Results are shown in table.~\ref{tab:overall_table_pub}. This dataset only contains user/item ID features and ratings, making it impossible to equip ADI with domain interest attention layer module~(Because only user/item ID feature is incorporated in the user/item tower). In addition, there are no overlap users between domains in this dataset shown in table~\ref{tab:dataset_overlap}. In our final implementation version, we use the backbone network and DSBN module. Therefore our ADI doesn't get an overall best performance on all domains but only achieves the best average performance.

\subsection{Discussions} \label{sec:discussion}
To better understand the effectiveness of our proposed ADI, We conduct several interesting discussions on the Alibaba Production dataset.

\begin{table}[htbp]
 \setlength{\abovecaptionskip}{0.5pt}
 \setlength{\belowcaptionskip}{0.5pt}
\caption{Variant versions' performances of the fusion layer.}\vspace{5pt}
\label{tab:fusion}

 \scalebox{0.8}{
    \begin{tabular}{c c c c c}
    \hline
    \hline
    \multirow{2}*{Variants} & \multicolumn{3}{c}{ R@500}\\
    \cmidrule(lr){2-4}&  Domain \#1 & Domain \#2 & Domain \#3 \\
    \hline 
    SUM & 0.4567 & 0.4105 & 0.2686 \\ 
    Network-Mul& 0.4619 & 0.4023 & 0.2706 \\
    \textbf{CONCAT} & \textbf{0.4683} & \textbf{0.4222} & \textbf{0.2982}  \\
    \hline
    \hline
    \end{tabular}
}
\end{table}
\subsubsection{Variants of the fusion layer}\label{sec:fusion_experiment}
As mentioned in Section~\ref{sec:fusion}, We implement different versions of the fusion layer. It is worth noting that our proposed method equiped with \textbf{Network-Mul} is exactly the same to the model architecture of STAR~\cite{sheng2021one} for the MDR problem. The results are showed in Table~\ref{tab:fusion}. Our proposed \textbf{CONCAT} achieves the best, while the original \textbf{SUM} version of MMoE/PLE~\cite{ma2018modeling, tang2020progressive} meets the worst performance. The performance of \textbf{Network-Mul} version used by STAR is between two variants.

\subsubsection{Ablation Study}

\begin{table}
 \setlength{\abovecaptionskip}{0.5pt}
 \setlength{\belowcaptionskip}{0.5pt}
\caption{Ablation study.}\vspace{5pt}
\label{tab:abaltion}
\resizebox{0.8\linewidth}{!}{
\begin{tabular}{c c c c c}
\hline
\hline
\multirow{2}*{Method} & \multicolumn{3}{c}{ R@500}\\
\cmidrule(lr){2-4}&  Domain \#1 & Domain \#2 & Domain \#3 \\
\hline 
~\textbf{ADI-SE} & \textbf{0.4683} & \textbf{0.4222} & \textbf{0.2982} \\
w/o ST & 0.4462 & 0.4022 & 0.2744 \\
w/o ST\&DIAL & 0.4164 & 0.3784 & 0.2576  \\
w/o ST\&DIAL\&DSBN & 0.3964 & 0.3555 & 0.2399 \\ 
\hline
\hline
\end{tabular}}
\end{table}

To study the effectiveness of each component of ADI, we conduct several ablation studies. We list those models without part of components as follows.
\begin{itemize}
\item {\textbf{ADI-SE}}: Full model version of ADI.
\item {\textbf{w/o ST}}: ADI trained without self training method.
\item {\textbf{w/o ST\&DIAL}}: ADI without domain interest adaptation layer and self training method.
\item {\textbf{w/o ST\&DIAL\&DSBN}}: ADI without DSBN, domain interest adaptation layer and self training method.
\end{itemize}
DSBN to capture the diversity of data distributions from different domains, domain interest adaptation layer to assign different feature weights for different domains, and self-training method to exploit potential label-level connections between domains, all improve the performance of the model as can be seen in Table~\ref{tab:abaltion}

\subsubsection{Impact on numbers of shared experts}
The numbers of shared expert networks effect the final model performance. Therefore, we change the shared expert networks of several methods from 1 to 5. As can be seen from Figure~\ref{fig:expert_num}, we compare MMoE, PLE and our proposed ADI with different numbers of shared expert networks in the exactly same model size setting~(For MMoE, all experts are shared experts).
The results show that with the shared expert networks increasing, the performances of PLE and ADI increase, while MMoE's increases at the beginning then decreases.

\begin{figure}[htbp]
  \includegraphics[width=0.5\textwidth]{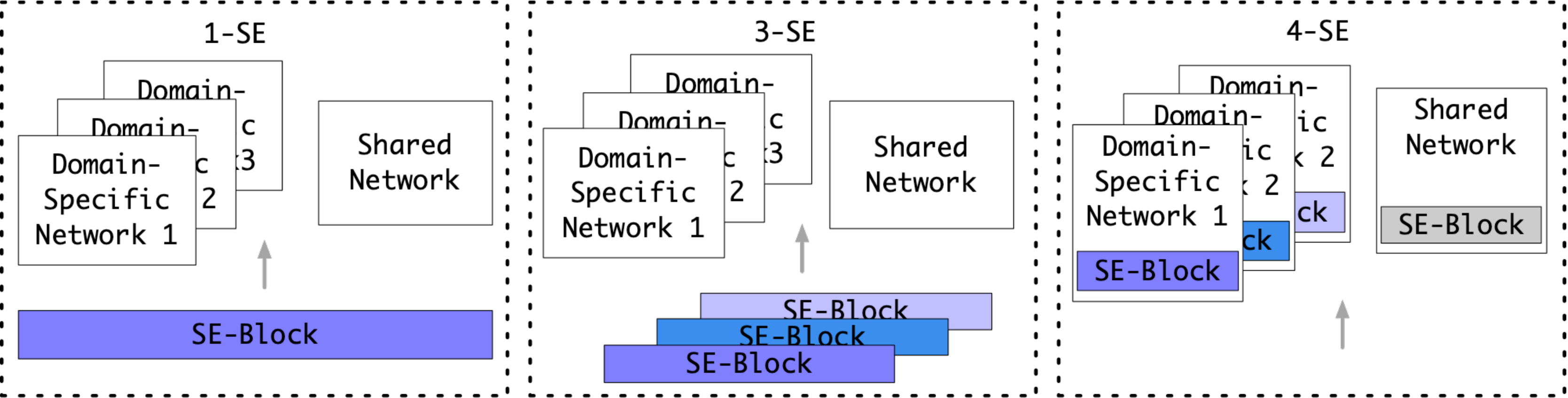}
  \setlength{\abovecaptionskip}{1pt}
 \setlength{\belowcaptionskip}{1pt}
  \caption{Different variants of SE-Block usages. 1-SE version denotes applying a global SE-Block to reweight raw features. 3-SE version denotes using the SE-Block selected by domain indicator to process feature-level domain adaptation. 4-SE version denotes equipping each domain-specific network and shared network with a SE-Block, in order to increasing the model capability.}
  \label{fig:se_usage}
\end{figure}

\subsubsection{Usage of SE-Block}
\begin{table}[htbp]
 \setlength{\abovecaptionskip}{1pt}
 \setlength{\belowcaptionskip}{1pt}
\caption{Variant versions' performances of SE-Blocks.}\vspace{5pt}
\label{tab:se_usage}
 \scalebox{0.8}{
\begin{tabular}{c c c c c}
\hline
\hline
\multirow{2}*{Variants} & \multicolumn{3}{c}{ R@500}\\
\cmidrule(lr){2-4}&  Domain \#1 & Domain \#2 & Domain \#3 \\
\hline 
1-SE & 0.4415 & 0.3922 & 0.2804 \\ 
\textbf{3-SE} & \textbf{0.4683} & \textbf{0.4222} & \textbf{0.2982} \\
4-SE & 0.4516 & 0.4041 & 0.2917 \\
\hline
\hline
\end{tabular}
}
\end{table}

Different variants of adding SE-Block into the backbone model gain differential improvements of the model performance. As shown in Figure~\ref{fig:se_usage}, we compare three usages of SE-Block, which are described as follows. Here we use the simplest version of ADI which consist 3 domain-specific networks and 1 shared network.
\begin{itemize}
\item {\textbf{1-SE}}: We use SE-Block as an universal domain interest adaptation layer for dynamically feature-level domain adapting.
\item {\textbf{3-SE}}: We keep $D$ domain adaptation layers with $D$ SE-Blocks. Which domain adaptation layer to use is depended on the domain indicator. 
\item {\textbf{4-SE}}: SE-Block is added into each expert network. In this variant SE-Block is treated as a method of increasing model complexity rather than domain adaptation.
\end{itemize}
Based on results shown in Table~\ref{tab:se_usage}, we argue that proper usage of SE-Block matters for the final model performance. The version of 1-SE gets the worst performance and 3-SE gets the best. An interesting observation is that 4-SE version is no better than 3-SE, even though the model parameters increase. Therefore, we suggest using \textbf{3-SE} version as the domain adaptation layer, which re-weight feature embeds at the sample domain-level rather than the network-level.

\subsubsection{Visualization}

\begin{figure}[htp]
 \setlength{\abovecaptionskip}{0.5pt}
 \setlength{\belowcaptionskip}{0.5pt}
  \includegraphics[width=0.5\textwidth]{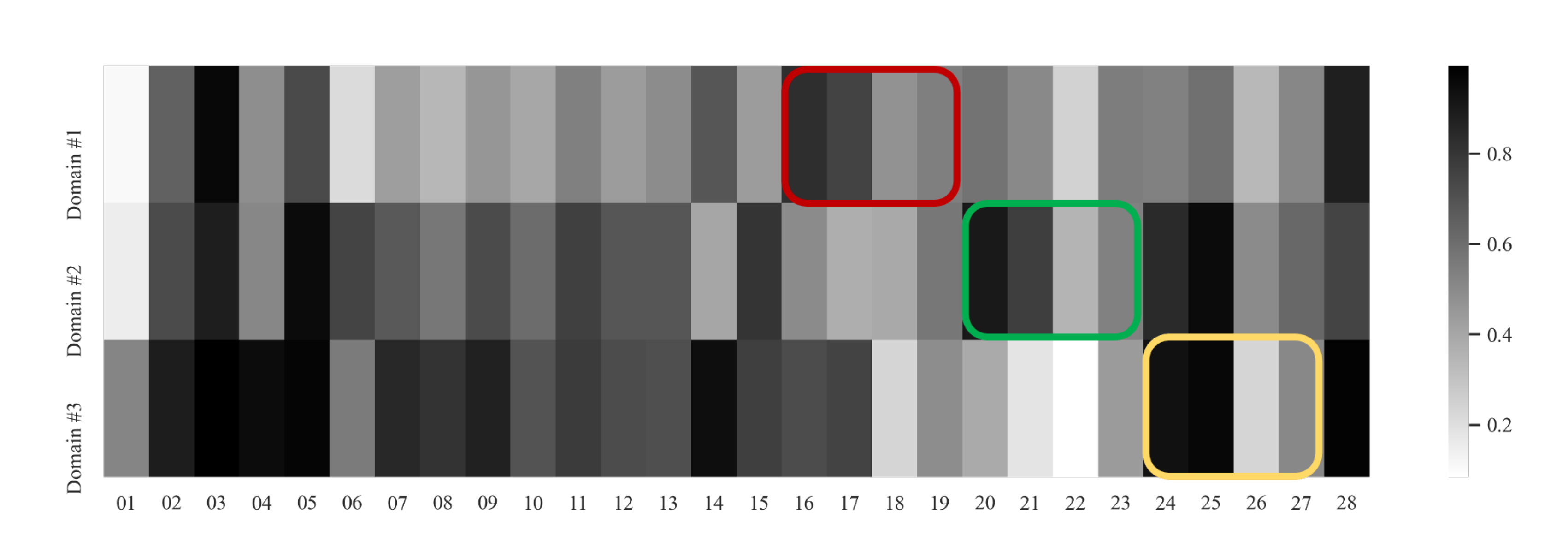}
  \caption{The attention weight visualization of each domain adaptation layer on item side.}
  \label{fig:se_visual}
\end{figure}

To give an intuitive evidence of effectiveness of the domain adaptation layer, we visualize the attention weights for different domains. As shown in Figure.~\ref{fig:se_visual}, the different darkness distributions suggest that each domain focuses on different features. It is worth note that we draw domain-related features in colors. To be more specific, we draw statistic features~( CTR, CLICK, COST and PAYNUM)~from domain \#1 in red, domain \#2 in green and domain \#3 in yellow. As we can seen, the adaptation layer automatically assign domain-related features higher attention scores, which proved the effectiveness of our proposed domain adaptation layer.

\begin{table}[htbp]
 \setlength{\abovecaptionskip}{0.5pt}
 \setlength{\belowcaptionskip}{0.5pt}
\caption{Online A/B Test}\vspace{5pt}
\label{tab:online}
\scalebox{0.8}{
    \begin{tabular}{c c c c}
    \hline
    \hline
    Online Metric &  Domain \#1 & Domain \#2 & Domain \#3 \\
    \hline 
    RPM & +1.9\% & +0.7\% & +2.6\% \\ 
    PPC & +2.0\% & -0.2\% & +1.6\%  \\
    CTR & +0.0\% & +0.9\% & +1.0\%  \\
    \hline
    \hline
    \end{tabular}
}
\end{table}
\section{Online A/B Test}
We have deployed our model on the display advertising system of Alibaba. To get a stable conclusion, we observe the online experiment for two weeks. Four common metrics in advertising system are used to measure the online performance: RPM(Revenue Per Mille), PPC(Pay Per Click), CTR(Click Through Rate) . As the result shown in Table~\ref{tab:online}, the present method ADI gets overall improvements on three domains in our online A/B test experiment.

\section{Conclusion}
In this paper, we have investigated the problem of multi-domain recommendation. Compared with exsiting works, our proposed ADI attempts to apply domain adaptation on retrieval stage of a recommendation system. The backbone network effectively learns commonalities and diversities for multiple domains. The DSBN component and the domain interest adaptation layer are applied for feature-level domain adaptation. And the self training method captures potential label-level connections across domains. Experiments on public and industrial datasets validate the superiority of our proposed method. Extensive discussions verify the effectiveness of the proposed method and conduct suggestions of proper usage of each component. The proposed ADI has been deployed on Alibaba display advertising system and gain significant profits, validating the commercial values of the proposed method.
\printbibliography

\end{document}